\documentclass[twocolumn,showpacs,preprintnumbers,amsmath,amssymb,prl,superscriptaddress]{revtex4}

\usepackage{graphicx}
\usepackage{dcolumn}
\usepackage{bm}
\usepackage{dcolumn}
\usepackage{amsmath}
\usepackage{amssymb}
\usepackage{bbold}

\begin{document}


\title{Spin Hall effect of alloys: Extrinsic and intrinsic contribution}
\author{Stephan Lowitzer}
\address{%
Department Chemie, Physikalische Chemie, Universit\"at M\"unchen, Butenandstr. 5-13, 81377 M\"unchen, Germany\\}
\author{Martin Gradhand}
\affiliation{Max-Planck-Institut f\"{u}r Mikrostrukturphysik, Weinberg 2, D-06120 Halle, Germany}
\affiliation{Institut f\"{u}r Physik, Martin-Luther-Universit\"{a}t Halle-Wittenberg, D-06099 Halle, Germany}
\author{Diemo K\"{o}dderitzsch}
 \email{dkopc@cup.uni-muenchen.de}
\affiliation{%
Department Chemie, Physikalische Chemie, Universit\"at M\"unchen, Butenandstr. 5-13, 81377 M\"unchen, Germany\\}
\author{Dmitry V. Fedorov}
\affiliation{%
Institut f\"{u}r Physik, Martin-Luther-Universit\"{a}t Halle-Wittenberg, D-06099 Halle, Germany}
 \author{Ingrid Mertig}
\affiliation{Max-Planck-Institut f\"{u}r Mikrostrukturphysik, Weinberg 2, D-06120 Halle, Germany}
\affiliation{Institut f\"{u}r Physik, Martin-Luther-Universit\"{a}t Halle-Wittenberg, D-06099 Halle, Germany}
 \author{Hubert Ebert}
\affiliation{%
Department Chemie, Physikalische Chemie, Universit\"at M\"unchen, Butenandstr. 5-13, 81377 M\"unchen, Germany\\}

\date{\today}

\begin{abstract}
A fully relativistic description of the spin-orbit induced spin Hall
 effect is presented that is based on Kubo's linear response
 formalism.  Using an appropriate operator for the spin current density
 a Kubo-St\v{r}eda-like equation for the spin Hall conductivity (SHC)
 is obtained.  An implementation using the Korringa-Kohn-Rostoker
 (KKR) band structure method in combination with the Coherent Potential
 Approximation (CPA) allow detailed investigations on various alloy
 systems. A decomposition of the SHC into intrinsic and extrinsic
 contributions is suggested. Accompanying calculations for the
 skew-scattering contribution of the SHC using the Boltzmann equation
 demonstrate the equivalence to the Kubo formalism in the dilute
 alloy regime and support the suggested decomposition scheme.
\end{abstract}

\pacs{71.15.Rf,72.25.Ba,75.76.+j,85.75.-d}
\keywords{Suggested keywords}
\maketitle

The emerging research field spintronics has been developed very
rapidly during the last years.  The reason for the broad interest in
this field is based on the close connection to fundamental scientific
questions as well as its impact on technology \cite{WAB01,AS09}. In
this context, the spin Hall effect (SHE) is one of the most promising
phenomena. It denotes the observation that a charge current flowing
through a solid is accompanied by a transversal spin current.  This
occurs even for non-magnetic solids as was demonstrated by experiments
on pure Pt
\cite{KOS07}.

Both the anomalous Hall effect (AHE) in ferromagnets and the SHE are
caused by the influence of spin-orbit coupling (SOC).  Accordingly,
their theoretical description is quite similar
\cite{FFH81,JNM02,NSO+10,SCN+04,YF05,GYN05,GMCN08,Guo09,TKN+08,GMN09,GFZM10a,GFZM10b}.
For ideal systems an intrinsic mechanism was identified that allows to
express the corresponding response function in terms of the Berry
curvature \cite{JNM02,SCN+04}.  On this basis {\it ab initio}
calculations for the intrinsic spin Hall conductivity (SHC) were
performed \cite{YF05,GYN05,GMCN08,Guo09}.  As for the AHE, the
additional extrinsic SHC in dilute and concentrated alloys is ascribed
to skew and side-jump scattering caused by SOC.  The role of these
mechanisms for the SHE has been studied so far primarily by model
calculations \cite{TKN+08,GMN09}.  First principle calculations for
the extrinsic SHC of dilute alloys on the basis of the Boltzmann
formalism that account for the skew scattering mechanism have been
performed only very recently \cite{GFZM10a,GFZM10b}.  However, a
complete description of intrinsic and extrinsic mechanisms giving rise
to the SHE applicable to ideal as well as alloy systems, as it is
presented below, was missing so far.  As pointed out by several
authors
\cite{QX05,JPDQ06}, a central issue
for such an approach is an adequate definition for the spin current
density operator that accounts for SOC.  This was supplied
recently by Vernes et al.\ \cite{VGW07} by starting from the
Bargmann-Wigner four-vector spin polarization operator ${\cal T}$
\cite{BW48}. Demanding that the spin polarization is connected with the
spin current density via a corresponding continuity equation an
explicit expression for the spin current operator was given.

An adequate formal basis for the discussion of the SHE in non-magnetic
metals is supplied by Kubo's linear response formalism that allows to
derive an expression for the spin Hall conductivity tensor. To avoid
any approximation when dealing with SOC the underlying electronic
structure is described in a fully relativistic way by the
four-component Dirac formalism
\cite{Ros61}. As for the ordinary electrical conductivity the
perturbation due to the external electric field is represented in
terms of the current density operator $\hat{\bf j}$. In its
relativistic form this operator is given by:
\begin{equation}
\label{eq:j}
\hat{\bf j}  = -  | e |\,  c \, {\bm \alpha} \; ,
\end{equation}
where ${\bm \alpha}$ is the vector-matrix of the standard Dirac matrices
$\alpha_i$ \cite{Ros61} and the other quantities have their usual
meaning. The response function to be considered for the SHE is the
spin current density.  Considering for the z-component of the spin
polarization vector the current density along the x direction the
corresponding operator is given by \cite{LKE10,VGW07,FN001}:
\begin{equation}
\hat{J}^{\rm z}_{\rm x}= |e| c \alpha_{\rm x}
\left(\beta \Sigma_{\rm z} -\frac{1}{mc}\gamma_5\hat{p}_{\rm z}  \right) \;,
\end{equation}%
where $\hat{p}_{\rm z}$ is the canonical momentum operator,
$\Sigma_{\rm z}$ is the $z$ component of the vector of the relativistic
spin matrix, $\beta$ and $\gamma_5$ are Dirac matrices \cite{Ros61}.

Adopting a single-particle description of the electronic structure in
terms of the retarded ($G^+$) and advanced ($G^-$) Green's function
and restricting to $T=0$~K, an explicit expression for the SHC is
obtained that is similar to the Kubo-St\v{r}eda equation for the
anomalous Hall conductivity $\sigma_{\rm xy}$ of ferromagnetic
systems.  Considering for the spin polarization along $\hat{z}$ its
current density along $\hat{x}$ due to an electric field along
$\hat{y}$, the SHC $\sigma_{\rm xy}^{\rm z}$ is given by \cite{Low10}:
\begin{align}
\nonumber
\sigma_{\rm xy}^{\rm z} =&\frac{\hbar}{2\pi N \Omega}
 {\rm Tr}\,\big\langle \hat{J}_{\rm x}^{\rm z} G^+ \hat{j}_{\rm y}  G^-\big\rangle_{\rm c}
\\
\label{eq:SHE} & +\frac{|e|}{4\pi i N \Omega} 
{\rm Tr}\,\big\langle (G^+-G^-)(\hat{r}_{\rm x}\hat{J}_{\rm y}^{\rm z} 
-\hat{r}_{\rm y}\hat{J}_{\rm x}^{\rm z}) \big\rangle_{\rm c} \; ,
\end{align} 
where terms containing products of the retarded (or advanced) Green's
functions have been dropped \cite{NSO+10}. Due to symmetry the last term
is site-diagonal for the cubic systems considered here. As furthermore
all system considered here are metallic \cite{TKN+08,NHK10} it has been
omitted as well.

The electronic Green's function occurring in Eq.~(\ref{eq:SHE}) for
the Fermi energy $E_{\rm F}$ can be evaluated in a very efficient way
by use of the relativistic version of the multiple scattering or
Korringa-Kohn-Rostoker (KKR) formalism \cite{Ebe00}. This approach is
applicable to ideal systems but can also be applied straightforwardly
to disordered alloys. For this case the brackets $\langle ...
\rangle_{\rm c}$ in Eq.~(\ref{eq:SHE}) imply a configurational average that is performed within
the Coherent Potential Approximation (CPA) \cite{But85}.  For this
purpose the KKR-CPA approach used for the electrical conductivity
tensor of alloys on the basis of the Kubo-St\v{r}eda equation
\cite{LKE10a} has been adapted to Eq.~(\ref{eq:SHE}).  In particular,
this approach accounts explicitly for the so-called {\it vertex
corrections} those represent the difference in the correlated and
uncorrelated configurational averages of the type $ \langle
\hat{J}_{\rm x}^{\rm z} G^+
\hat{j}_{\rm y} G^-
\rangle_{\rm c}$ and $ \langle \hat{J}_{\rm x}^{\rm z} G^+ \rangle_{\rm c}
\langle \hat{j}_{\rm y} G^- \rangle_{\rm c} $, respectively. 
For the following it is important to note that the vertex corrections
correspond to the {\it scattering-in} term within the Boltzmann formalism
\cite{But85}.

Representing the anomalous Hall conductivity (AHC) $\sigma_{\rm xy}$
in terms of Feynman diagrams it was demonstrated that all extrinsic
contributions to $\sigma_{\rm xy}$ due to skew and side-jump
scattering correspond to terms involving vertex corrections
\cite{CB01a}. Obviously, the same conclusion can be drawn for the SHC
$\sigma_{\rm xy}^{\rm z}$. The remaining diagrams, involving no vertex
 corrections, stand for the intrinsic anomalous or spin Hall
 conductivity, plus corrections to this due to chemical disorder. It
 therefore seems natural to extend the definition of the intrinsic SHC
 $\sigma_{\rm xy}^{{\rm z}\,{\rm intr}}$ to diluted and concentrated
 alloys to represent all contributions not connected to the
 vertex-corrections. According to this definition, Eq.~(\ref{eq:SHE}),
 including the vertex corrections, gives the total SHC $\sigma_{\rm
 xy}^{\rm z} \equiv \sigma_{\rm xy}^{{\rm z}\,{\rm VC}} $ while the
 intrinsic SHC $\sigma_{\rm xy}^{{\rm z}\,{\rm intr}} \equiv
 \sigma_{\rm xy}^{{\rm z}\,{\rm no VC}} $ is obtained if those are
 ignored.  Thus, $\sigma_{xy}^{{\rm z}\,{\rm intr}}$ is the intrinsic
 SHC of the effective CPA medium specific for each composition of a
 certain alloy.  As a consequence, the extrinsic SHC $\sigma_{\rm
 xy}^{{\rm z}\,{\rm extr}} $ to be ascribed to the skew and side-jump
 scattering mechanisms is obtained from the difference $\sigma_{\rm
 xy}^{{\rm z}\,{\rm extr}}=\sigma_{\rm xy}^{\rm z} -
\sigma_{\rm xy}^{{\rm z}\,{\rm intr}}$.

\begin{figure}
 \begin{center}
 \includegraphics[width=8cm,clip]{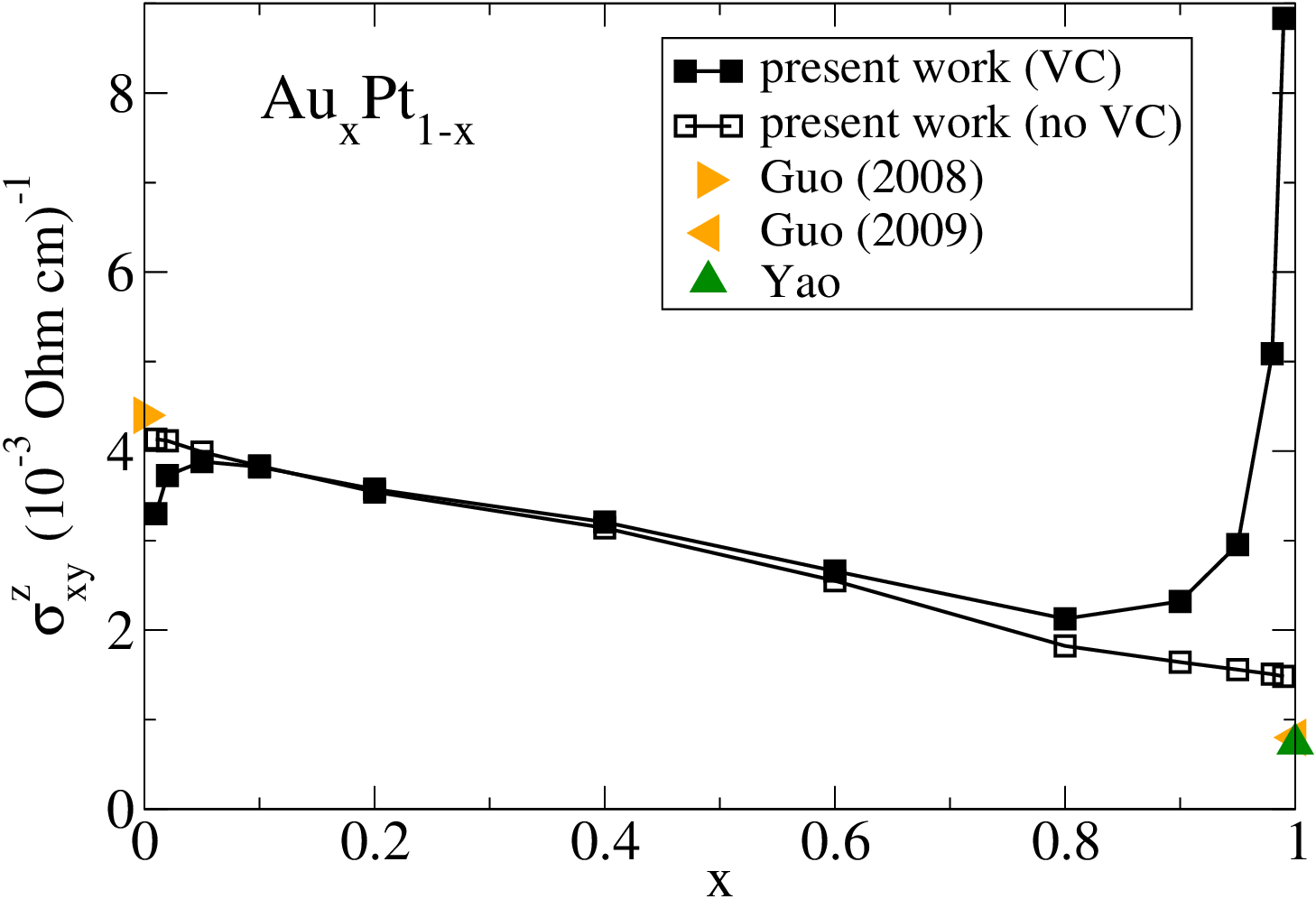}
 \vspace*{0.3cm}
 \includegraphics[width=8cm,clip]{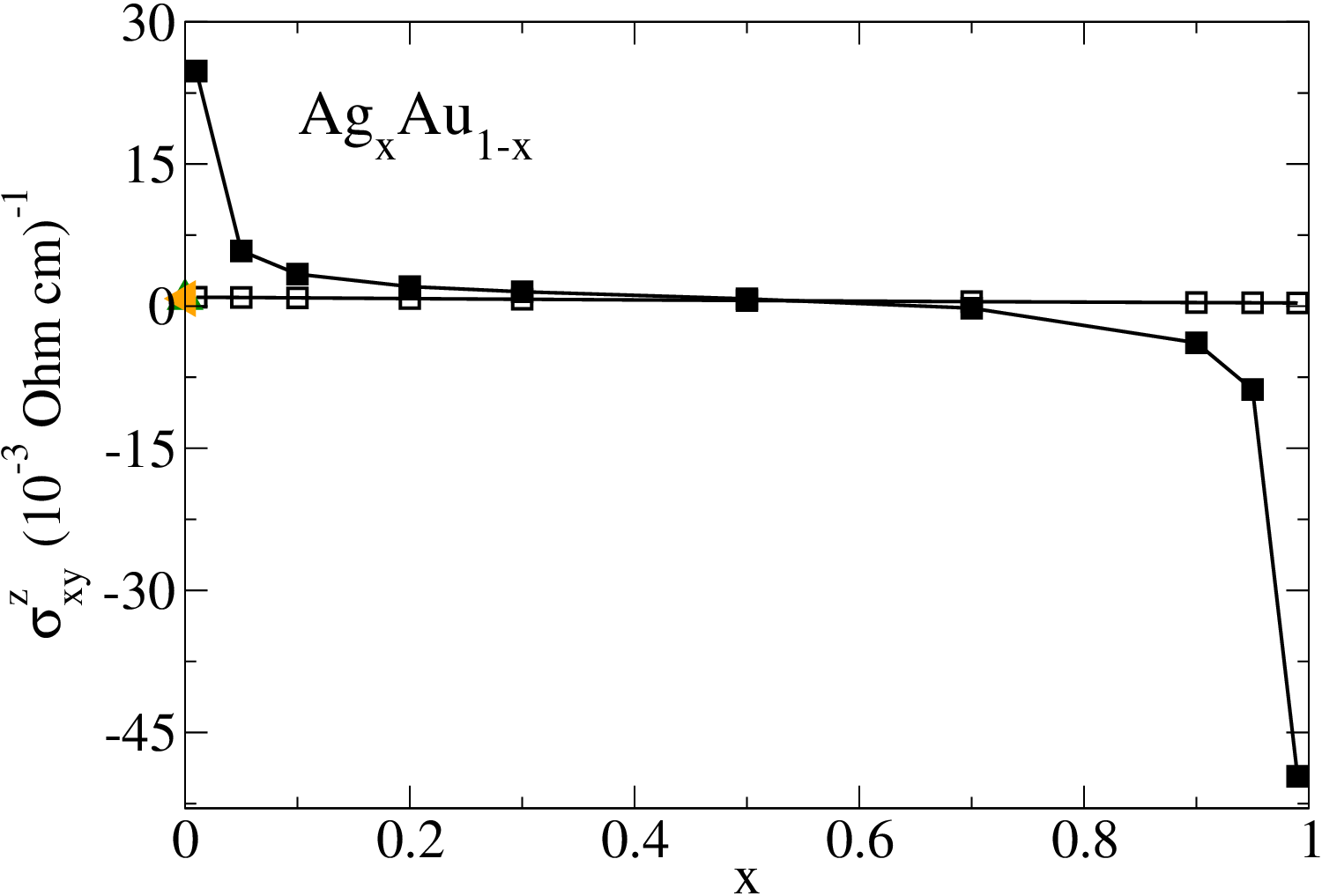} 
\caption{\label{plot:SHE}(Color online) Spin Hall
 conductivity $\sigma_{\rm xy}^{\rm z}$ for the alloy systems
 Au$_x$Pt$_{1-x}$ and Ag$_x$Au$_{1-x}$. The full squares correspond to
 calculations including vertex corrections while the open squares
 represent calculations without vertex corrections. Results from other
 $ab$ $initio$ investigations on the intrinsic SHE of pure Pt
 \cite{GMCN08} and Au \cite{YF05,Guo09} are included.  Because of the
 different definitions for the spin current operator the results from
 Refs.~\onlinecite{GMCN08} and 
\onlinecite{Guo09} have been multiplied by a factor of 2
 for the sake of consistency.}  \end{center}
\end{figure}

The approach sketched above has been applied 
to investigate the SHE for the fcc alloy systems 
Au$_x$Pt$_{1-x}$ and Ag$_x$Au$_{1-x}$. 
As can be seen from Fig.~\ref{plot:SHE} $\sigma_{\rm xy}^{{\rm z}\,{\rm intr}}$
 obtained from Eq.~(\ref{eq:SHE}) 
 ignoring the vertex corrections varies nearly linearly with the 
concentration throughout the whole composition regime. 
%
%
In addition, Fig.~\ref{plot:SHE} (top) shows results for the intrinsic
SHC of Au \cite{YF05,Guo09} and Pt \cite{GMCN08} obtained by other
authors using an expression for $\sigma_{\rm xy}^{{\rm z}\,{\rm
intr}}$ in terms of the Berry curvature. Taking into account the
differences between the various calculation schemes used -- in
particular concerning the treatment of spin-orbit coupling and the
definition of the spin current density operator -- these data fit
reasonably well with the alloy data obtained using the
Kubo-St\v{r}eda-like equation (Eq.~(\ref{eq:SHE})).  This obviously
justifies the extension of the definition for
$\sigma_{\rm xy}^{{\rm z}\,{\rm intr}}$ to the alloy case to represent
all terms that do not involve the vertex corrections.  

In contrast to the intrinsic SHC, the total one ($\sigma_{\rm
xy}^{{\rm z}}$) shows for both investigated alloy systems a divergent
behavior in the dilute limit when the concentration $x$ approaches 0
or 1, respectively. Interestingly, the corresponding extrinsic SHC
$\sigma_{\rm xy}^{{\rm z} \,{\rm extr}}$ changes sign when the
concentration $x$ varies from 0 to 1.  For concentrated alloys ($0.2<
x < 0.8 $) the intrinsic and total SHC do not differ strongly. As this
behavior is also found for other alloy systems that do not show a
change in sign for the extrinsic SHC $\sigma_{\rm xy}^{{\rm z} \,{\rm
extr}}$ it seems that the impact of the vertex corrections in the
concentrated alloy regime is in general negligible.

 For the anomalous Hall effect it is known that one may classify solid
state materials according to the scaling relation between the AHC
$\sigma_{\rm xy}$ and the longitudinal conductivity $\sigma_{\rm xx}$.
For metallic systems those fall in general into the so-called
ultra-clean regime ($\sigma_{\rm xx}
\gtrsim 10^6 \,(\Omega {\rm cm})^{-1}$) the skew-scattering mechanism
should dominate $\sigma_{\rm xy}$
\cite{OSN06,OSN08}.  In this case the scaling relation $\sigma_{\rm xy}\approx
\sigma_{\rm xy}^{\rm skew}  = S \,   \sigma_{\rm xx} 
$ holds with $S$ the so-called {\it skewness factor}.  Assuming the
same to apply for the SHE as well, one may expect for the extrinsic SHC
the relation:
\begin{equation}
\label{eq:scaling}
\sigma_{\rm xy}^{{\rm z} \,{\rm extr}} =
\sigma_{\rm xy}^{{\rm z} \,{\rm skew }} 
+
\sigma_{\rm xy}^{{\rm z} \,{\rm sj}}
=
S^{\rm z} \, \sigma_{\rm xx}
+
\sigma_{\rm xy}^{{\rm z} \,{\rm sj}} \; ,
\end{equation}
 where $S^{\rm z}$ is the corresponding skewness factor and the term
 $\sigma_{\rm xy}^{{\rm z} \,{\rm sj}}$ represents the contribution
 due to the side-jump mechanism.

\begin{figure}
 \begin{center}
 \includegraphics[width=8cm,clip]{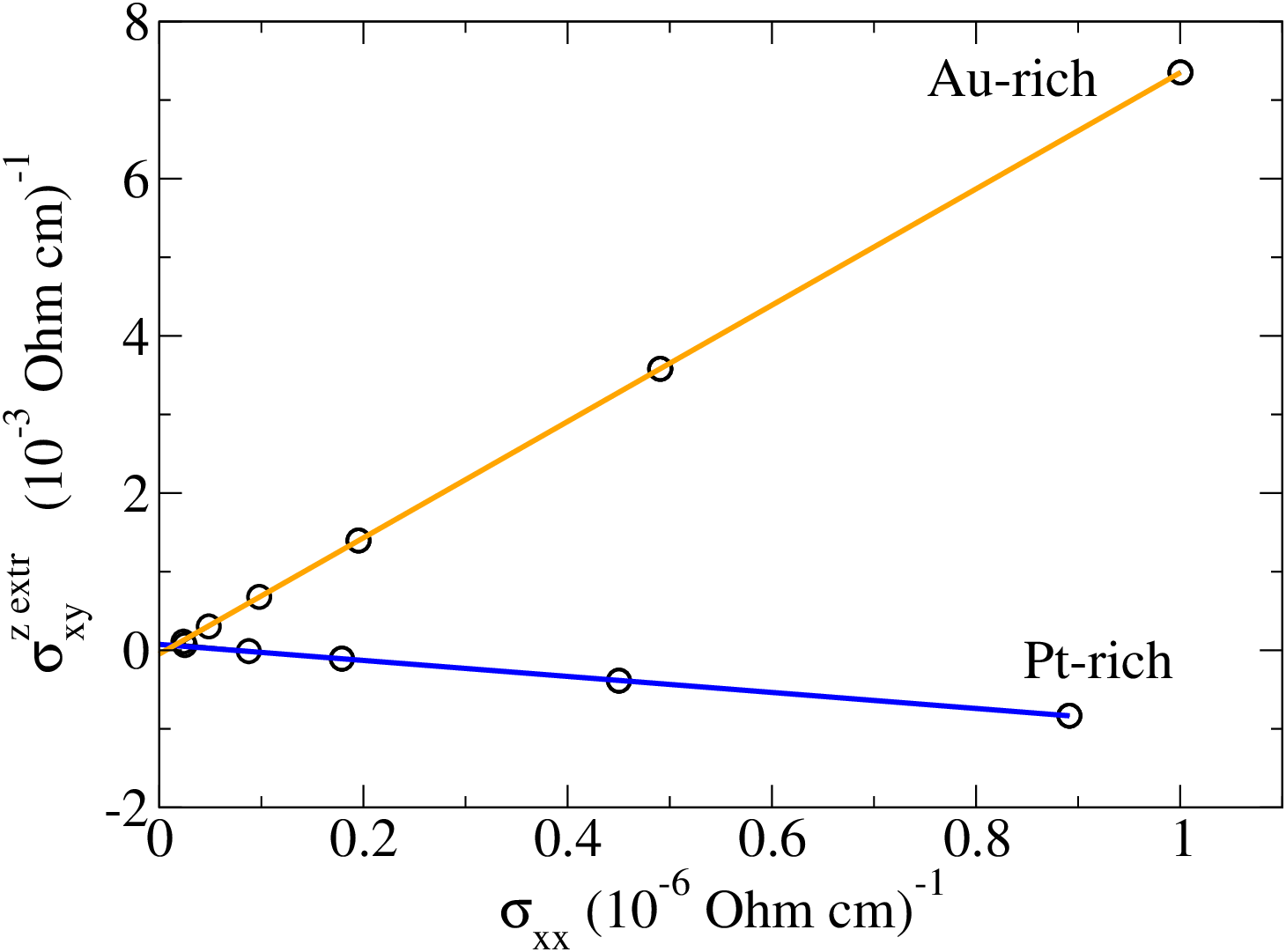}
 \vspace*{0.3cm}
 \includegraphics[width=8cm,clip]{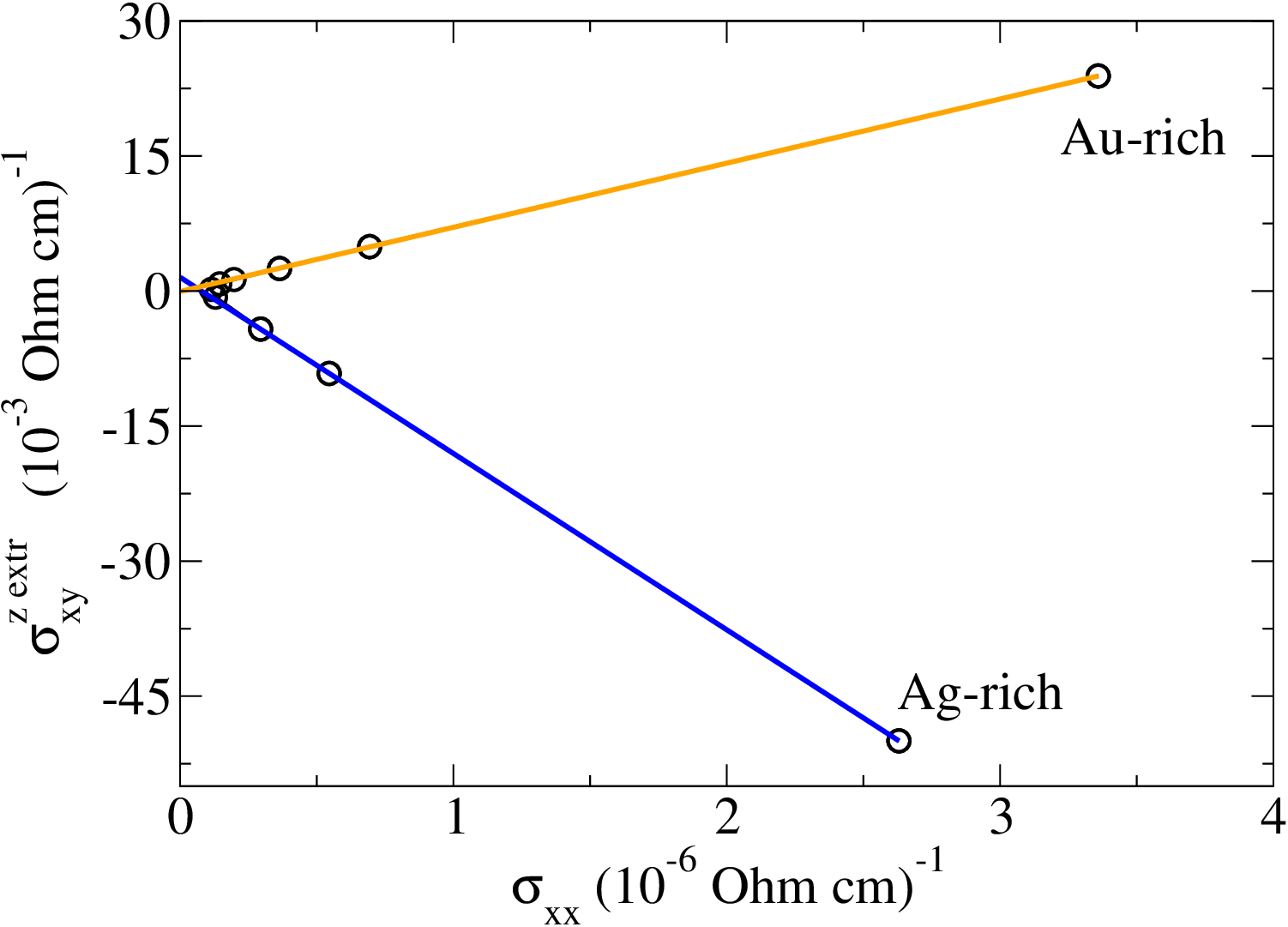} 
\caption{\label{plot:SHE_xy-vs-S_xx}(Color online)
 The extrinsic spin Hall conductivity $\sigma_{\rm xy}^{{\rm z} \,{\rm extr}}$ 
versus the longitudinal conductivity $\sigma_{\rm xx}$ 
for Ag$_x$Au$_{1-x}$ and Au$_x$Pt$_{1-x}$ (black line/circles). 
The blue and orange lines are explained in the text.}
 \end{center}
\end{figure}

\begin{figure}  
\begin{center}
 \includegraphics[scale=0.32,clip]{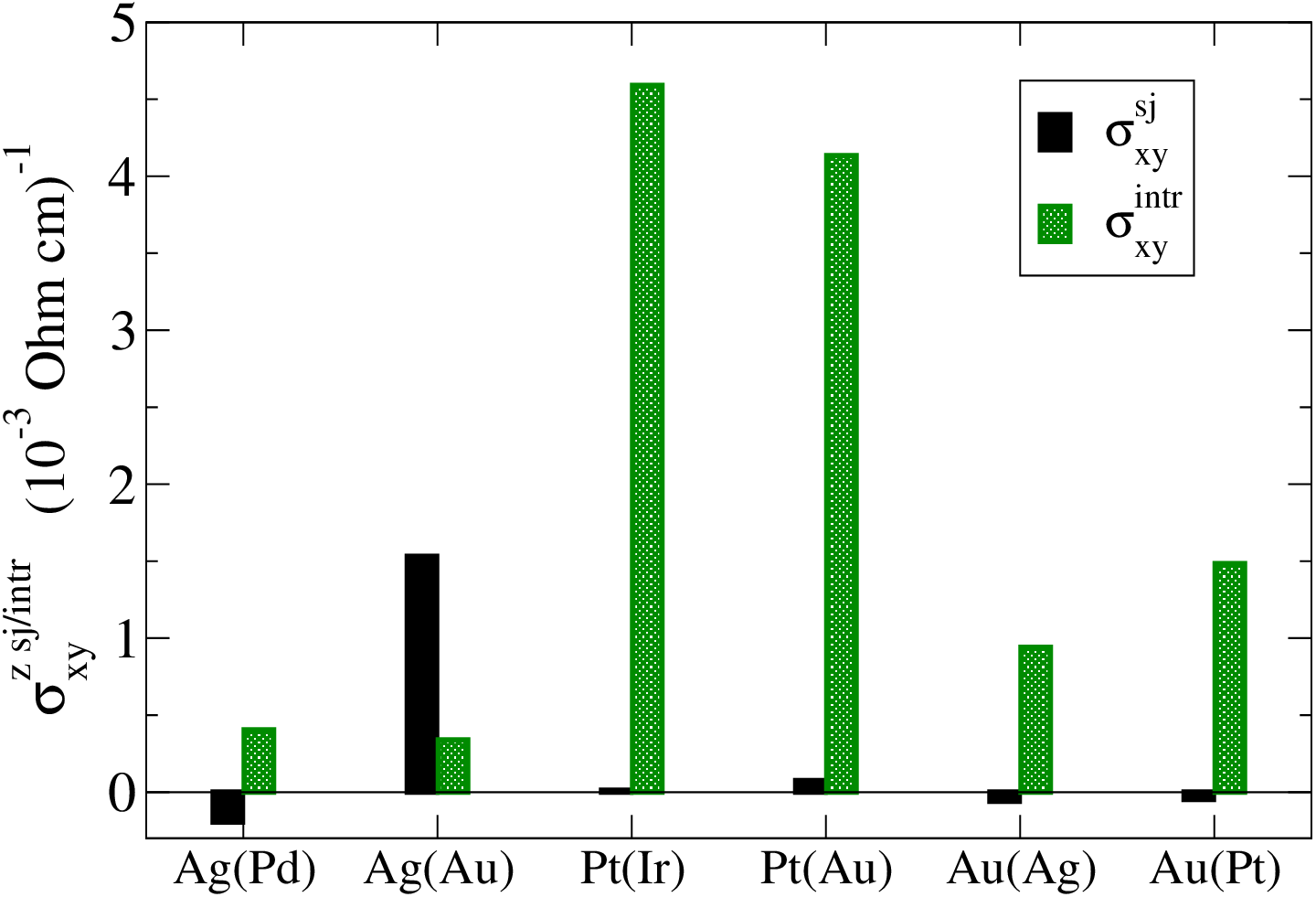}
 \caption{\label{plot:sxx_side_jump}(Color online) 
The side-jump and intrinsic contributions to the spin Hall conductivity,
 $\sigma_{\rm xy}^{{\rm z}\,{\rm sj}}$ 
and $\sigma_{\rm xy}^{{\rm z}\,{\rm intr}}$,
 respectively, for various dilute transition
 metal alloys A(B) with the  concentration of 1 at.\% 
for the dissolved component B.}
\end{center}
\end{figure}

Plotting the extrinsic SHC $\sigma_{\rm xy}^{{\rm z}\,{\rm extr}} $ of
 Au$_x$Pt$_{1-x}$ and Ag$_x$Au$_{1-x}$ versus the corresponding
 $\sigma_{\rm xx}$ with the concentration as
 an implicit parameter indeed a linear behavior is found in the dilute
 regimes ($x\leq 0.1$ or $x\geq 0.9$), as can be seen from
 Fig.~\ref{plot:SHE_xy-vs-S_xx}.
%
%
Fitting a straight line to the data for the considered systems and extrapolating to $\sigma_{\rm xx}=0$ allows to deduce the side-jump contribution on the basis of Eq.~(\ref{eq:scaling}). The
results obtained for Au$_x$Pt$_{1-x}$ and Ag$_x$Au$_{1-x}$ are shown in
Fig.~\ref{plot:sxx_side_jump} together with data obtained for two
other alloy systems.
%
%
Obviously, $\sigma_{\rm xy}^{{\rm z}\,{\rm sj}} $ may take either sign
and is in most cases found to be much smaller than the intrinsic
contribution.  On the other hand, $\sigma_{\rm xy}^{{\rm z}\,{\rm
skew}} $ is dominating in the dilute limit since it scales with the
concentration, while $\sigma_{\rm xy}^{{\rm z}\, {\rm intr}}$ does not
depend on the concentration explicitly.  However, an actual impurity
concentration for the crossover between intrinsic and skew scattering
regime depends on the considered alloy.

\begin{figure}
 \begin{center}
 \includegraphics[width=8cm,clip]{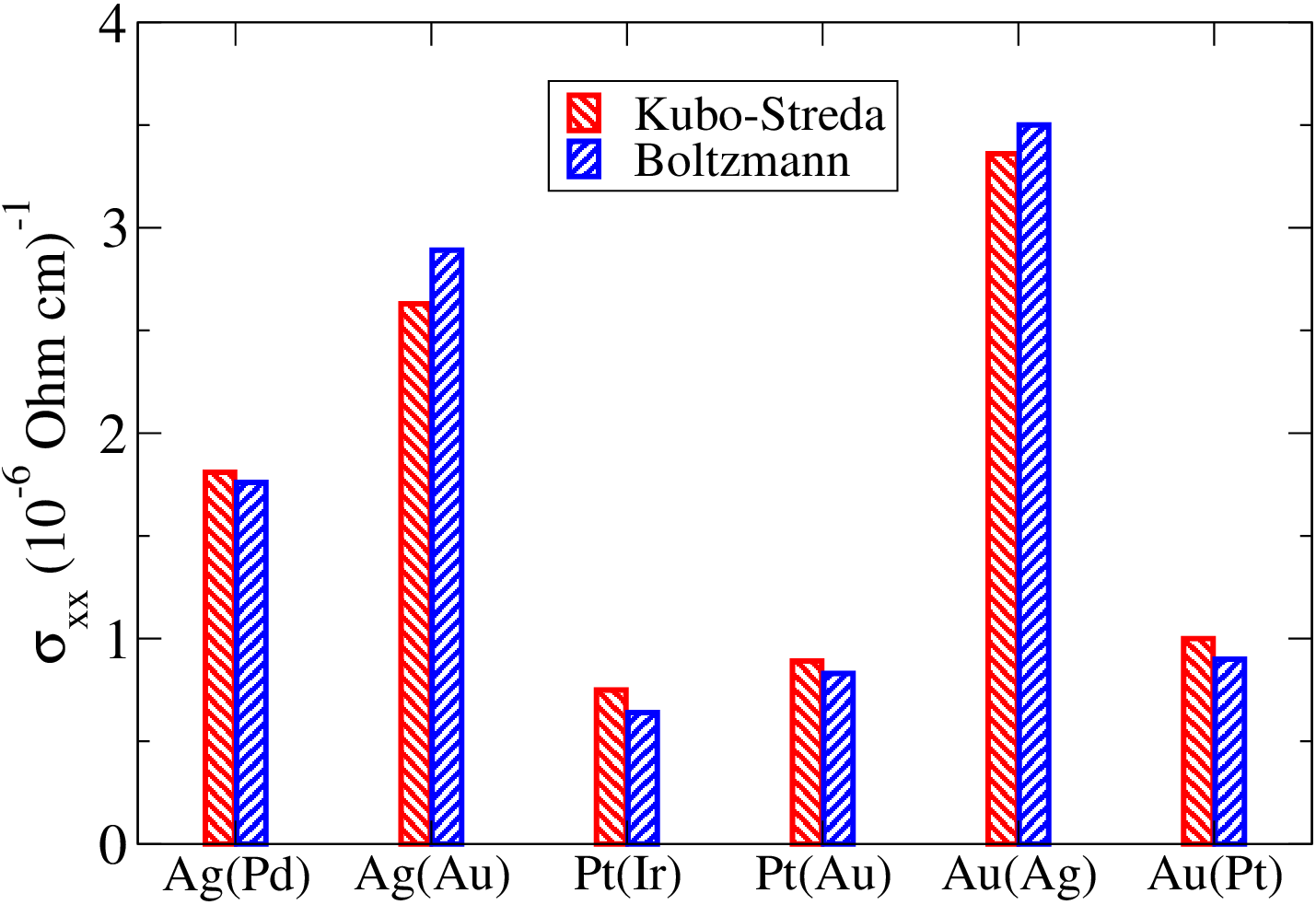}
 \vspace*{0.3cm}
 \includegraphics[width=8cm,clip]{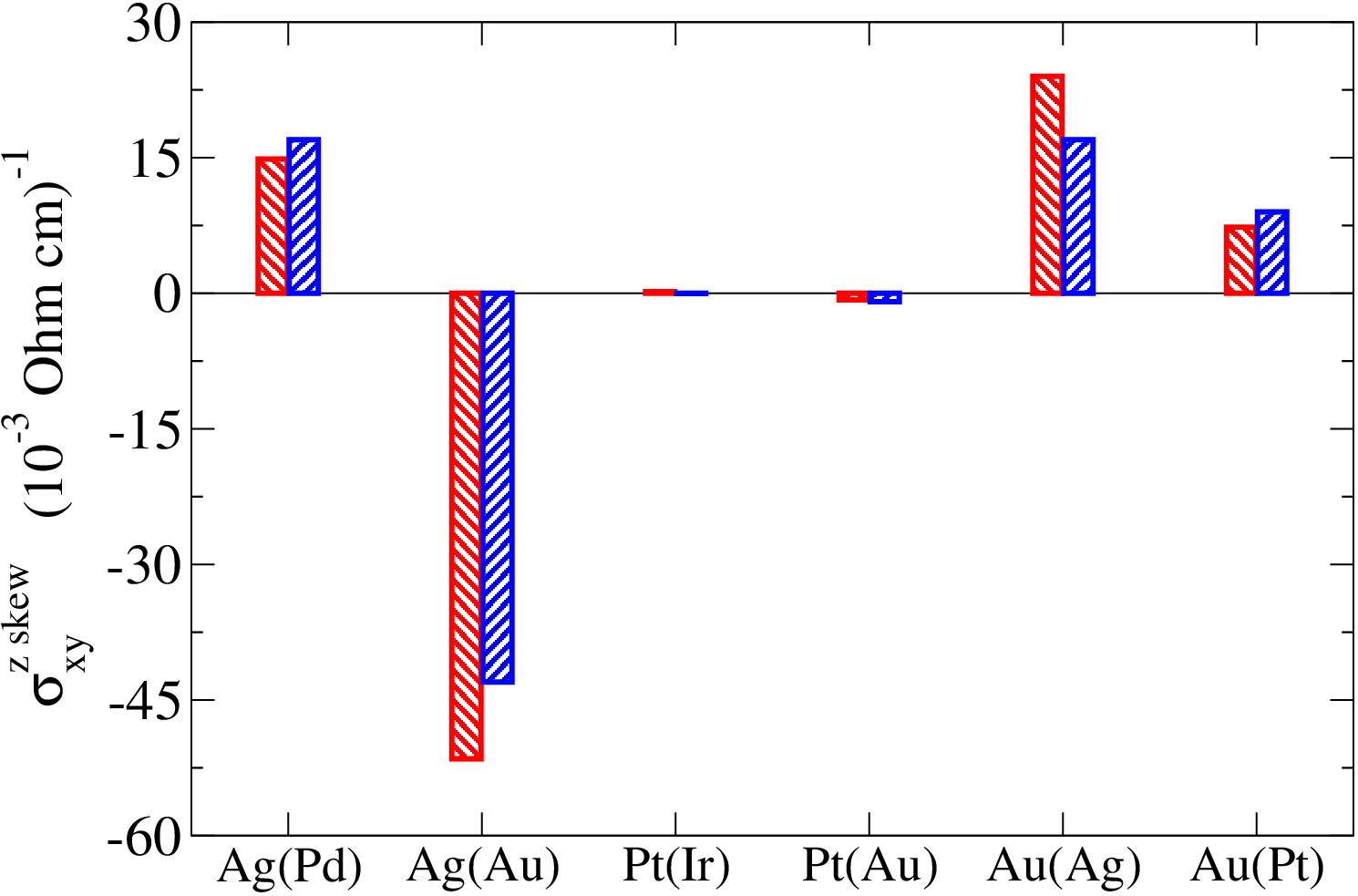} 
\caption{\label{plot:sxx}(Color online) 
The longitudinal conductivity $\sigma_{\rm xx}$ (top) and the 
skew scattering contribution to the spin Hall conductivity 
$\sigma_{\rm xy}^{{\rm z}\,{\rm skew}}$ (bottom) for various 
dilute alloys A(B) obtained on the basis of the Kubo-St\v{r}eda 
equation (left bar) and the Boltzmann formalism (right bar), 
respectively. The concentration for the dissolved component B is 1 at.\%. }
 \end{center}
\end{figure}

 To support the analysis of the results for the total SHC presented
above, complementary work has been done using the Boltzmann formalism for the SHE
\cite{GFZM10a}. The results for $\sigma_{\rm
xx}$ obtained this way are found in very good agreement with those
obtained using the Kubo-Greenwood equation (see top panel of
Fig.~\ref{plot:sxx}).
%
%
 As mentioned above, the vertex corrections giving rise to
 $\sigma_{\rm xy}^{{\rm z}\,{\rm extr}}$ correspond to the
 scattering-in processes occurring in the Boltzmann formalism. As it
 was demonstrated recently, the latter ones give rise to the
 skew-scattering mechanism \cite{GFZM10a}.  Corresponding results for
 $\sigma_{\rm xy}^{{\rm z}\,{\rm skew}} $ (bottom panel of
 Fig.~(\ref{plot:sxx})) are also found in very satisfying agreement
 with the results based on the Kubo-St\v{r}eda-like formula
 (Eq.~(\ref{eq:SHE})) together with the described decomposition.  This
 finding convincingly shows the equivalence of both approaches for the
 dilute alloy regime and it also justifies once more the used
 definition for the intrinsic SHC introduced above.

In summary, an expression for the spin Hall conductivity $\sigma_{\rm
 xy}^{\rm z}$ has been derived in analogy to the Kubo-St\v{r}eda
 equation for the anomalous Hall conductivity $\sigma_{\rm xy}$ of
 ferromagnets. An implementation within the fully relativistic KKR-CPA
 formalism allows numerical investigations for various transition
 metal alloy systems over the whole range of composition.  We
 decompose the total SHC $\sigma_{\rm xy}^{{\rm z}}$ into its
 intrinsic and extrinsic parts.  In the concentrated alloys the
 intrinsic contribution of the effective medium always dominates. The
 extrinsic contribution, on the other hand, shows in general a
 diverging behavior in the dilute alloy regime that is ascribed to the
 skew scattering contribution.  Accompanying calculations on the basis
 of the Boltzmann formalism demonstrate its equivalence with the Kubo
 formalism in the dilute alloy regime and support the decomposition
 made for the total SHC $\sigma_{\rm xy}^{{\rm z}}$.

The authors S.L., D.K. and H.E. would like to thank the DFG for
financial support within the SFB 689 ``Spinph\"anomene in reduzierten
Dimensionen''. Further more, this work was supported by the
International Max Planck Research School for Science and Technology.

\begin{thebibliography}{27}
\bibitem[{\citenamefont{x}(2010)}]{FN001}
\bibinfo{note}{Note, that different conventions are used in the literature 
 for the spin-current. For the sake of comparison with literature we       
 construct $\hat{J}^{\rm z}_{\rm x}$ such that it is given in units of     
 charge conductivity. In addition, it has to be mentioned that the spin    
 current was defined with opposite sign in previous publications by        
 some of the authors \cite{GFZM10a,GFZM10b}.}
\end{thebibliography}

\end{document}